\documentclass[prl,aps,reprint,showpacs,showkeys,floats,floatfix,superscriptaddress, citeautoscript]{revtex4-2}
\usepackage{graphicx}
\usepackage{dcolumn}
\usepackage[colorlinks,allcolors=black,citecolor=blue,urlcolor=blue]{hyperref}
\usepackage{comment}

\begin{document}

\title{
Aromatic Molecule Solvation in Liquid Water with Coupled Cluster Accuracy:\\The Balance of Pi-Interactions and Hydrophobicity
}

\author{Nore Stolte}
\email{nore.stolte@theochem.rub.de}
\affiliation{Lehrstuhl f{\"u}r Theoretische Chemie, Ruhr-Universit{\"a}t Bochum, 44780 Bochum, Germany}
\author{Harald Forbert}
\affiliation{Center for Solvation Science ZEMOS, Ruhr-Universit{\"a}t Bochum, 44780 Bochum, Germany}
\author{Yury Lysogorskiy}
\affiliation{Interdisciplinary Centre for Advanced Materials Simulation (ICAMS), Ruhr-Universit{\"a}t Bochum, 
44780
Bochum, Germany}
\author{Ralf Drautz}
\affiliation{Interdisciplinary Centre for Advanced Materials Simulation (ICAMS), Ruhr-Universit{\"a}t Bochum, 
44780
Bochum, Germany}
\author{Dominik Marx}
\affiliation{Lehrstuhl f{\"u}r Theoretische Chemie, Ruhr-Universit{\"a}t Bochum, 44780 Bochum, Germany}


\begin{abstract}
Aromatic organic solutes in water exhibit a delicate balance between hydrophobic solvation and directional O--H$\cdots \pi$ hydrogen bonds, yet widely used force fields and 
state-of-the-art
density functional approaches struggle to provide a consistent picture of these 
pivotal 
interactions.
We introduce a data-efficient upfitting strategy to train a machine learning interatomic potential
(MLIP) based on the graph atomic cluster expansion
for aqueous
aromatic molecules with CCSD(T) accuracy
for condensed phase simulations,
using only finite molecular clusters.
We apply our method to aqueous toluene (C$_6$H$_5$CH$_3$).
The resulting CCSD(T)‑quality MLIP reproduces coupled cluster energies and forces in bulk and reveals
that commonly employed 
methods
do not capture the 
crucial
balance between hydrophilic and hydrophobic solvation, distorting the interactions of aromatic molecules with their environment.
Representative biomolecular force fields substantially understructure the hydrophobic solvation shell and misorient interfacial water, while overestimating $\pi$-contacts, yielding an inconsistent solvation balance.
Even hybrid DFT and MP2 overestimate barriers to breaking of water--$\pi$ hydrogen bonds.
Our workflow provides a 
practical,
general route to CCSD(T)-quality condensed-phase simulations of aqueous solutions,
and thus constructed 
interaction potentials
now
open the door to consistent, highly accurate 
benchmark
studies of $\pi$-contacts and hydrophobic effects in biomolecular contexts
such as solvation of 
proteins and DNA
in aqueous environments. 
\end{abstract}

\maketitle

\section*{Introduction}

Aromatic groups are found in proteins, nucleic acids, small-molecule ligands, 
and many environmental contaminants.
Their interactions with the aqueous environment
are central determinants of structure, intermolecular bonding, and molecular recognition.
The
intermolecular interactions via the delocalized $\pi$-electrons in the aromatic ring systems are well-known to
affect 
biomolecular function such as 
DNA double helix stabilization, 
protein folding, 
DNA--protein and 
protein--ligand binding and recognition 
as well as enzyme reaction mechanisms~\cite{Steiner2001Hydrogen, Meyer2003Interactions, Salonen2011Aromatic, Riley2013Importance}.
Further, many environmental pollutants include an aromatic group, and these molecules 
while dissolved in water
pose significant environmental challenges and complicate remediation efforts~\cite{Ossai2020Remediation, Dutta2024Contamination, Tang2026Research}.
Although interactions of organic molecules with water are typically classified as hydrophobic,
the 
extended 
$\pi$-orbital
systems 
of aromatic groups
accept H-bonds from water molecules~\cite{Suzuki1992Benzene, Tsuzuki2000Origin, Arunan2011Defining, Arunan2011Definition, Gierszal2011Hydrogen, Choudhary2019Abinitio, 
Chen2023Elucidating},
thereby
impacting hydrophobic solvation~\cite{Schravendijk2005Hydrophobic}.
Given their ubiquity and broad implications, understanding the properties of 
aromatic compounds in aqueous solution 
is of vital importance to, among others, drug design, biotoxicity screening, and industrial waste-water treatment approaches.

It is known that non-covalent intermolecular interactions involving $\pi$-electrons of aromatic molecules are not well-reproduced by many
popular force fields~\cite{Chessari2002Evaluation, Allesch2008First, Sherrill2009Assessment, Paton2009Hydrogen, Fu2011Comparative, Garrido2011Predicting}
which are,
nevertheless, used in a myriad of biomolecular simulation studies. 
For example, only after custom corrections to the atomic partial charges did 
force field simulations of alkyl-aromatic compounds in water 
reproduce experimental results~\cite{Garrido2011Predicting}.
When it comes to \textit{ab initio} molecular dynamics simulations~\cite{MarxHutter2009} (AIMD),
density 
functional theory (DFT) 
does
not appear to systematically correct for the discrepancies between force fields and experiment~\mbox{\cite{Paton2009Hydrogen, Prampolini2015Accuracy, Ajala2019Assessment}}.
Polarizable force fields are able to capture, to some extent, the anisotropy in interaction strength of aromatic molecules~\cite{Stone1988Some, Lemkul2016Empirical, Zhang2018AMOEBA}, albeit at increased computational cost.
A recent endeavor to include atomic-level anisotropy into all force field terms provides 
improvement of the radial structure of liquid benzene relative to experiment, but falls short of offering a 
broader agreement~\cite{Janicki2023Development}.

To resolve this fundamental problem,
it
has been suggested that calculations at quantum-chemical CCSD(T) accuracy are required in order to correctly describe aromatic interactions with atomistic detail~\cite{Riley2013Importance}. 
For 
sufficiently small
finite systems, e.g., benzene 
or similar aromatic compounds that are 
microsolvated by a few water molecules, 
such 
static 
correlated wave function calculations are accessible
since a long time~\mbox{\cite{Kim2000Molecular, Chen2023Elucidating}}.
Yet, bulk solutions might display significantly different solvation properties to finite clusters: In the liquid state, there is competition between water--water and water--solute interactions, which determines both the hydrophobic solvation shell structure and the extent of 
H-bonding between water and the $\pi$-electron cloud. 
This phenomenon has been found to generate far-ranging modulations of the solvent structure
and dynamics
around hydrophobic groups~\cite{Imoto2018Aqueous, ContiNibali2020Wrapping, DasMahanta2023Local}.
Despite such urgent need,
correlated wave function calculations for aromatic molecules in bulk liquid water 
at the required CCSD(T) accuracy
have not been feasible
so far.

\begin{figure}[b]
\includegraphics[width=8.6cm]{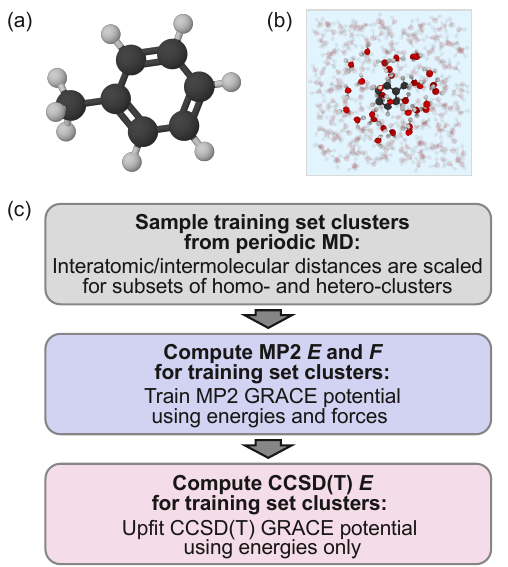}
\caption{Scheme illustrating the training procedure of the CCSD(T) GRACE 
interaction potential
for aqueous toluene.
(a) The toluene molecule, showing one of the resonance structures.
(b) Toluene in a periodic simulation cell with 256 water molecules. The 32 water molecules with the smallest C--O distances, contributing 
toluene-water hetero-cluster configurations
to the training set, are highlighted.
(c) Training scheme of the CCSD(T) GRACE interatomic potential;
see 
Fig.~\ref{fig:scheme_extended}
for further details.
}
\label{fig:scheme}
\end{figure}

To achieve a wave-function-based description of bulk aqueous solutions
at CCSD(T) accuracy
represents
an outstanding challenge.
Machine learning interatomic potentials (MLIPs) present a way to circumvent the expensive electronic structure calculations needed during molecular dynamics simulations, by training the MLIP to the level of theory needed for the simulation;
this is now routinely done using periodic training sets from AIMD in conjunction with DFT electronic structure. 
Beyond DFT, MLIPs have been trained to 
CCSD(T)
data to provide an accurate description of bulk liquid water~\cite{Daru2022Coupled, Chen2023Data-Efficient, ONeill2025Towards},
providing excellent agreement with 
many experimental properties including
isotopic substitution effects \cite{Stolte2024Nuclear, Paschek2025When, Stolte2025Scrutinizing}.
A recent work explores how to achieve a 
CCSD(T) description
that is localized in space
for aqueous ion pairs
by
combining density functional embedding theory with machine learning~\cite{Bian2026Transfer}.
Additionally, there are efforts 
that complement MLIP-based approaches
to promote the simulation accuracy of bulk water~\cite{Yu2023Status, Palos2024Current}
and aqueous solutions~\cite{Zhou2025Water, Boittier2025Roadmap, Wang2026Cluster} to coupled cluster theory using various approaches 
including many-body expansion techniques
to parameterize the potential energy surface.
Explicit 
periodic
coupled cluster molecular dynamics simulations
of water
are simply inaccessible due to the scaling of the calculation in combination with the large 
periodic supercell
sizes needed to describe 
the 
dynamically disordered H-bonded liquid.
In stark contrast, MLIPs offer~-- 
in principle~-- 
a pathway to studying bulk aqueous systems with coupled cluster accuracy.

Moving beyond pure liquid water, in particular to address the solvation of 
large molecular species such as aromatic organic compounds, 
adds significant complexity when aiming to train MLIPs at CCSD(T) accuracy. 
A key challenge to their construction relates to the reference structures 
that make up the training set, for which
coupled cluster calculations must be performed. 
These structures must capture the molecular variety of solvation patterns
encountered in 
solutions as well as the cross-over to the bulk liquid, 
which seems to require very large solute--solvent complexes.
But even
with coupled cluster implementations with favorable scaling~\cite{Riplinger2013Efficient, Liakos2015Possible, Neese2019Chemistry}, 
the realistically accessible system sizes
are limited to relatively small, 
finite molecular complexes,
with on the order of hundreds of atoms.

\begin{figure*}
\includegraphics[width=17.2cm]{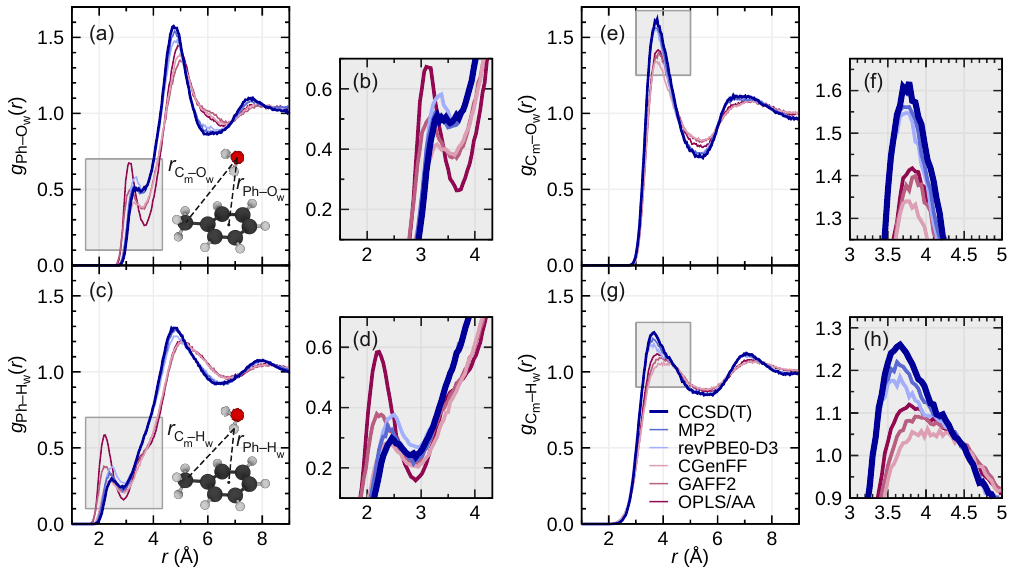}
\caption{Radial structure of aqueous toluene 
from simulations with classical nuclei using the CCSD(T), MP2, and revPBE0-D3 GRACE interatomic potentials, and CGenFF, GAFF2, and OPLS/AA force fields.
The RDFs are computed with respect to the center of mass of the carbon atoms in the ring, denoted Ph, and the methyl carbon atom, denoted C$_{\rm m}$.
Water oxygen atoms and hydrogen atoms are referred to as O$_{\rm w}$ and H$_{\rm w}$, respectively.
Insets in (a) and (e) illustrate the relevant coordinates.
(a)~Ph--O$_{\rm w}$ RDF. (b)~Zoom of the shaded area in~(a).
(c)~Ph--H$_{\rm w}$ RDF. (d)~Zoom of the shaded area in~(c).
(e)~C$_{\rm m}$--O$_{\rm w}$ RDF. (f)~Zoom of the shaded area in~(e).
(g)~C$_{\rm m}$--H$_{\rm w}$ RDF. (h)~Zoom of the shaded area in~(g).
The RDFs use a binwidth of 0.04~{\AA}, and curves are not smoothed.
}
\label{fig:rdf}
\end{figure*}

In this work, we have developed a strategy to 
systematically
construct 
an exhaustively diverse
training set for an MLIP with CCSD(T) accuracy for the simulation of
aromatic molecules in liquid water
at ambient conditions,
using 
exclusively
finite clusters of molecules that together describe the inhomogeneous system,
in conjunction with a suitable two-step approach to reach CCSD(T) accuracy. 
We used the graph atomic cluster expansion (GRACE) to parameterize the interatomic many-body interactions~\cite{Bochkarev2024Graph},
and employed upfitting 
\cite{Lysogorskiy2026Graph}
to reach the final CCSD(T) accuracy.
The entire procedure is built on correlated wave function theory, utilizing a lower-level model trained to MP2 energies and forces, and an upfitted model trained to CCSD(T) energies,
thus avoiding the use of DFT~training data.
We demonstrate in Sections~S4 and S5 in the Supporting Information (SI) that the quality of CCSD(T) forces is independent of the quality of the forces provided by the lower-level model; 
hence 
upfitting from either a DFT or MP2 reference leads to
CCSD(T) accuracy for both energies and forces in the bulk while exclusively using finite cluster training data. 
A summary of 
our 
training procedure is 
sketched 
in Fig.~\ref{fig:scheme}(c)
while 
Fig.~\ref{fig:scheme_extended}
unfolds details.

We 
employ the general procedure that we developed here 
to train a CCSD(T) interaction potential for aqueous toluene 
(C$_6$H$_5$CH$_3$,
also known as methylbenzene or toluol,
Fig.~\ref{fig:scheme}(a)).
As a widely-used commodity chemical, 
toluene
is a volatile organic compound commonly used 
in chemical synthesis, 
as an important ``BTEX compound" in 
gasoline, 
and as
solvent in coatings and paints~\cite{Ziegler-Sylakakis2019Toluene}.
In aqueous solution, the hydration structure of toluene is affected both by hydrophobic interactions of the benzene ring and the methyl group, as well as the 
directional 
\mbox{O--H$\cdots \pi$}
hydrogen bonds with water molecules which involve the aromatic $\pi$-system. 
Therefore,
aqueous
toluene 
is
a prototypical molecule with both 
a bulky
aliphatic, hydrophobic group and an aromatic 
$\pi$-system,
both of which are
ubiquitously 
present 
for instance 
in proteins 
and DNA/RNA~nucleobases.
For
these reasons, we consider toluene to be the archetype 
molecule
to investigate
\textit{in~nuce} the role of 
non-covalent $\pi$-bonds 
and hydrophobic interactions 
of biomolecules with aqueous environments. 
Our 
MLIP reproduces CCSD(T) accuracy for bulk aqueous toluene at ambient conditions, and is used to study the subtle 
hydrophobic and $\pi$-interactions 
involved in the hydration of toluene, 
including 
the impact of
nuclear quantum effects.

\section*{Results}

\subsection*{Toluene solvation structure}
We computed radial distribution functions (RDFs) of water oxygen and hydrogen atoms, denoted O$_{\rm w}$ and H$_{\rm w}$, respectively, with respect to the methyl carbon atom of toluene (C$_{\rm m}$) and the center of mass of the carbon atoms in the aromatic ring, denoted Ph (for phenyl), as a geometric proxy for the center of the $\pi$-system of toluene.
The RDFs that probe water molecules near the $\pi$-system reveal that all 
interaction potentials considered 
qualitatively
predict H-bonding of water to the $\pi$-system (Fig.~\ref{fig:rdf}(a--d)),
i.e., \mbox{$\pi$-hydrogen bonds}.
Thus, the 
\mbox{Ph--H$_{\rm w}$} RDF has a peak at shorter distances than the \mbox{Ph--O$_{\rm w}$} RDF, 
indicating that water molecules are oriented with an \mbox{O$_{\rm w}$--H$_{\rm w}$} bond pointing 
with H$_{\rm w}$
towards the center of the phenyl ring in toluene.
Despite this apparent success of 
force field molecular dynamics (FFMD) 
and DFT simulations to describe $\pi$-water interactions,
the height of the first peak in the Ph--H$_{\rm w}$ RDF 
and also the depth of the subsequent minimum
is found to vary
greatly between the different potentials 
with reference to the CCSD(T) benchmark~--
with no clear trend across density functional and force field simulations.
This is expected to lead to sizable differences in the
free energy barrier to break and make O--H$\cdots \pi$ hydrogen bonds
being critical to the solvation of otherwise hydrophobic aromatic species in water,
which we will analyze in detail below.

\begin{figure*}
\includegraphics[width=17.2cm]{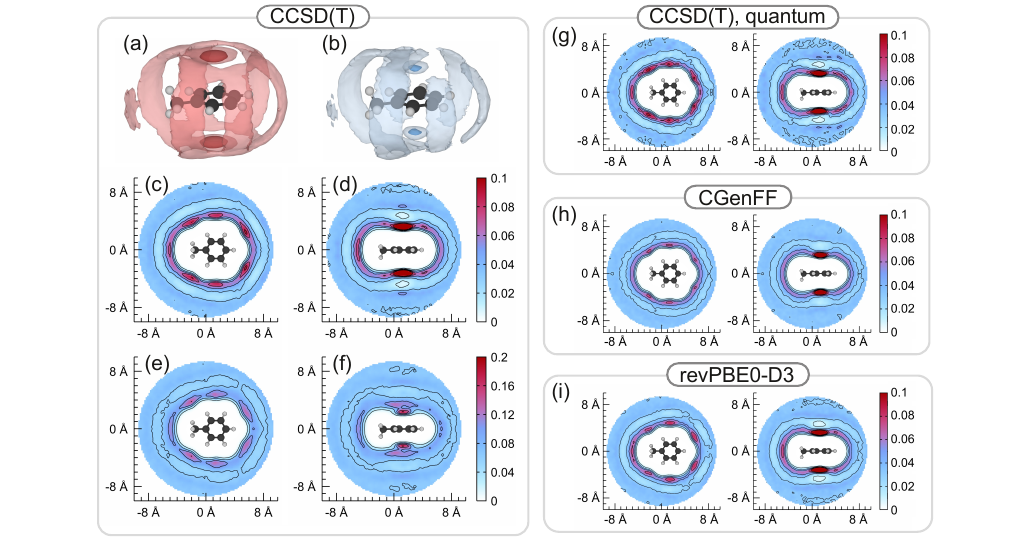}
\caption{Solvation structure of aqueous toluene.
(a) Spatial distribution function (SDF) of O$_{\rm w}$ atoms near toluene. Isosurfaces are drawn at \mbox{0.13~{\AA}$^{-3}$} (opaque) and \mbox{0.07~{\AA}$^{-3}$} (transparent). 
(b) SDF of H$_{\rm w}$ atoms near toluene. Isosurfaces are drawn at \mbox{0.16~{\AA}$^{-3}$} (opaque) and \mbox{0.11~{\AA}$^{-3}$} (transparent).
(c) Slice of the O$_{\rm w}$ SDF that contains the plane of the phenyl group.
(d) Slice of the O$_{\rm w}$ SDF that is perpendicular to the phenyl group and contains the methyl carbon atom, in {\AA}$^{-3}$.
Contour lines are drawn at intervals of \mbox{0.02~{\AA}$^{-3}$} in the range \mbox{0.01--0.1~{\AA}$^{-3}$}.
(e,f) Slices of the H$_{\rm w}$ SDF as in (c,d), in~{\AA}$^{-3}$.
Contour lines are drawn at intervals of \mbox{0.04~{\AA}$^{-3}$} in the range \mbox{0.02--0.2~{\AA}$^{-3}$}.
Results shown in (a--f) are obtained with the CCSD(T) GRACE interaction potential and classical nuclei.
\mbox{(g,h,i)} Slices of the O$_{\rm w}$ SDFs as in (c,d), from simulations using respectively the CCSD(T) GRACE interaction potential with quantum nuclei, 
the CGenFF force field with classical nuclei, and
the revPBE0-D3 GRACE interaction potential with classical nuclei.
The SDFs were computed w.r.t. carbon atoms of toluene, and were symmetrized w.r.t. the two symmetry planes of the carbon backbone.
In these figures, hydrogen atoms are placed to guide the eye.
The SDFs use a voxel side length of 0.2~{\AA}, and were smoothed using Gaussian smoothing with $\sigma = 0.16$~{\AA} in all three dimensions. 
}
\label{fig:solvationshell}
\end{figure*}

Scrutinizing next the hydrophobic solvation of the aliphatic group,
we turn to the RDFs involving the methyl carbon atom (C$_{\rm m}$)
in Fig.~\ref{fig:rdf}(e--h). 
In these, there is no distinct first peak that indicates H-bonding, and the first peak positions of the \mbox{C$_{\rm m}$--O$_{\rm w}$} and \mbox{C$_{\rm m}$--H$_{\rm w}$} RDFs are nearly the same.
As in the RDFs involving the aromatic ring center, the predictions from the different FFMD and 
machine learning molecular dynamics (MLMD) 
simulations vary 
significantly~-- but differently from the $\pi$-water contacts. 
The \mbox{revPBE0-D3} DFT functional describes the hydrophobic interactions better w.r.t. CCSD(T),
comparable to MP2, while it deviates at the same time
considerably for the $\pi$-interactions as we have seen. 
The force fields now all agree in significantly underestimating the first peak of 
the \mbox{C$_{\rm m}$--O$_{\rm w}$} and \mbox{C$_{\rm m}$--H$_{\rm w}$} RDFs
relative to the CCSD(T) benchmark, and thereby underrate interactions 
of this compound with water due to hydrophobic solvation.

Overall, having now access to converged reference data at 
quantum-chemical CCSD(T) accuracy as provided by our GRACE MLMD simulations, we
must conclude that these force fields and 
also the advanced dispersion-corrected 
hybrid DFT interatomic potential 
all fail to provide a consistent description of the key interactions of this generic aromatic compound with water,
namely hydrophobic solvation and intermolecular $\pi$-interactions.

In the discussion so far, we have compared the radial structure of aqueous toluene from simulations with classical nuclei.
To obtain quantitative predictions,
we used the CCSD(T) GRACE interatomic potential in a path integral simulation,
which samples the quantum mechanical partition function of the nuclei and thus includes nuclear quantum effects.
Our simulations reveal that 
nuclear quantum effects are negligible for the solvation structure of toluene
(Fig.~S18 in the SI).
Even the H-bonding peak in the Ph--H$_{\rm w}$ RDF is not visually different between the classical and the path integral simulation, which is surprising on account of the relatively large extent of delocalization of light H atoms in water, especially in H-bonds~\cite{Stolte2024Nuclear}.
The H-bond acceptor, being the $\pi$-system of toluene, is so localized that the ensemble of donated H atoms is no more delocalized than in the simulation with classical nuclei.
Hence,
this comparison justifies our analysis of simulations with classical nuclei to understand the variation in solvation structure of aromatic molecules predicted using several different 
interaction potentials
and their comparison to the CCSD(T) benchmark.

Going beyond radial structure, the
three-dimensional solvation structure of toluene
is now
probed
by 
symmetry-adapted
spatial distribution functions (SDFs)
obtained from CCSD(T) GRACE simulations.
The SDF involves
six lobes of higher water density in the equatorial region of the phenyl ring, centered 
not at but rather
between the toluene hydrogen atoms (Fig.~\ref{fig:solvationshell}(a--f); see Section S8 in the SI for SDFs from all simulations).
It 
qualitatively
mirrors the solvation structure of aqueous benzene found from AIMD with the PBE functional~\cite{Allesch2007Structure}.
For toluene, 
that water cage around the aromatic ring
is disrupted by the hydrophobic methyl group, near which there is a slight increase in the water density directly in line with the C--C bond relative to the rest of the first solvation shell.
At the same time, and 
supporting now
previous findings for aqueous benzene~\cite{Allesch2007Structure, Choudhary2015Spatial}
at CCSD(T) accuracy, 
there is a density maximum for both oxygen and hydrogen above and below the phenyl ring, in the region where $\pi$-interactions play a role.
The toluene solvation structure qualitatively reproduces that of benzene near the aromatic group, 
but is distorted by the steric and hydrophobic effects of the methyl group on the other side of the molecule.

How are these competing solvation effects described by force fields and 
hybrid
DFT?
The toluene solvation shell from the revPBE0-D3 GRACE interaction potential is understructured everywhere, except in the region of $\pi$-contacts.
We also found understructuring beyond the first solvation shell by the revPBE0-D3 GRACE potential in the RDFs, and from the SDFs, it is clear that
in particular in the plane of the 
phenyl ring,
\mbox{revPBE0-D3} provides
less structuring of the liquid with reference to the CCSD(T) benchmark (Fig.~\ref{fig:solvationshell}(c) and (i)).
Of the force fields under consideration, CGenFF describes best the 
collective interactions of aromatic carbon atoms with water,
rather close to the CCSD(T) description of the $\pi$-interactions
between water and toluene.
At the same time, CGenFF fails to provide a 
consistently accurate description of those intermolecular contacts where 
water molecules are facing only one or two carbon atoms.
Similar results regarding understructuring of the solvation shell near hydrophobic regions of toluene are found for the OPLS/AA and GAFF2 force fields, 
while these interaction potentials 
now 
overstructure regions where water interacts with the $\pi$-electrons.

In summary, we find that the hybrid DFT GRACE potential and 
force fields that represent three distinct families 
all
provide significantly deviating results for the toluene solvation shell relative to CCSD(T) accuracy.
Most critically, these interaction potentials are consistently unable to resolve the issue of hydrophobic 
solvation
versus $\pi$--water 
interactions of aromatic species in aqueous solution
as probed by toluene.

\subsection*{Hydrophobic solvation pattern}

\begin{figure}[b]
\includegraphics[width=8.6cm]{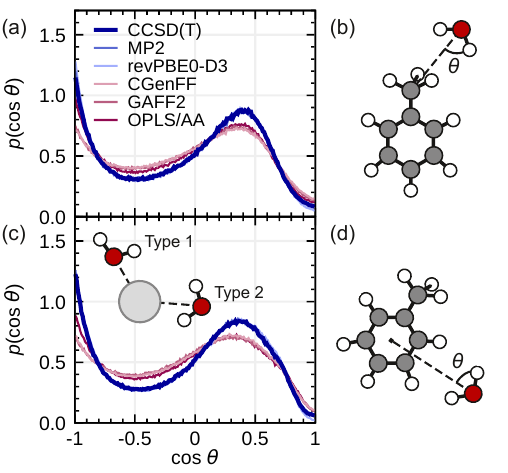}
\caption{Probability distributions of cosine of straddling angles $\theta_{\rm C_m \cdots O_w - H_w}$ (a,b) and $\theta_{\rm Ph \cdots O_w - H_w}$ (c,d) of solvating hydrophobic water (see text).
Water molecules that were determined to be H-bonded to the toluene $\pi$-orbital
system
(Fig.~\ref{fig:hbonds})
were excluded from the straddling angle analysis. 
}
\label{fig:hydrophobic}
\end{figure}

We have seen that
there are rather large 
differences
in the first solvation shell of the methyl group depending on the interaction potential.
The CGenFF model underestimates hydrophobic solvation shell structuring the most severely (Fig.~\ref{fig:rdf}), followed by GAFF2 and then OPLS/AA.
In contrast, 
the $\pi$-contacts are best described by CGenFF 
whereas OPLS/AA now fails badly, 
which adds serious inconsistencies to the overall balance of the toluene--water interactions. 
The orientation of water molecules solvating nonpolar solutes that do not engage in 
H-bonding
has been found to consist of ``straddling" configurations~\cite{Stillinger1973Structure, Geiger1979Molecular, Stillinger1980Water}, where three of the four H-bond directions per water molecule straddle the solute, 
while
the fourth one points away from the solute.
This leads to two distinct types of H$_2$O configurations, either with an \mbox{O$_{\rm w}$--H$_{\rm w}$} bond pointing away from the solute (type 1), or with both \mbox{O$_{\rm w}$--H$_{\rm w}$} bonds straddling the solute (type 2) (see inset in Fig.~\ref{fig:hydrophobic}(c)).
We analyzed the 
water molecules near toluene to uncover 
how the different potentials describe the straddling phenomenon compared to the CCSD(T) benchmark. 
The 
interfacial
water molecules 
were determined 
independently from their \mbox{H-bonding} pattern by
using the GITIM procedure~\cite{Sega2013Generalized}
(as implemented in the Pytim package~\cite{Sega2018Pytim}, based on water oxygen atoms with a Van der Waals radius of 1.58~{\AA}, and a probe sphere radius of 1.8~{\AA}).
Then, 
interfacial
water molecules were determined to be solvating either the methyl group or the phenyl ring, based on the \mbox{O$_{\rm w}$--Ph} and \mbox{O$_{\rm w}$--C$_{\rm m}$} distances: If $d_{\rm O_w-Ph} < d_{\rm O_w-C_m}$, then the molecule solvated the phenyl group, and \textit{vice versa};
water 
molecules that donated an H-bond to toluene
(see below)
were excluded from the analysis.

The three electronic structure-based GRACE potentials predict nearly identical water orientations near toluene (Fig.~\ref{fig:hydrophobic}), with a peak in the straddling angles in both populations 
around $\theta \approx 65^{\circ}$ ($\cos \theta \approx 0.4$) and for \mbox{$\cos = 180^{\circ}$} (\mbox{$\cos \theta = -1.0$});
these
straddling angles are consistent with the illustrated configurations.
The three force fields under consideration have less well-defined peaks, and the distribution of orientations of hydrophobic solvating water molecules is thus more broad, with only a slight preference for the straddling configurations.
The weaker interactions between water and hydrophobic regions of the toluene molecule provided by the three force fields provide fewer restrictions on the orientations of water molecules in the first solvation shell than
provided by the CCSD(T) benchmark.
Not only is the hydrophobic solvation shell density underestimated with these force fields, but the water orientations near the hydrophobic regions are also not correctly represented, so that overall, the solvation structure of aromatic organic molecules 
predicted
by these popular methods is unreliable.

\subsection*{Hydrophilic $\pi$-water interactions}

\begin{figure}[b]
\includegraphics{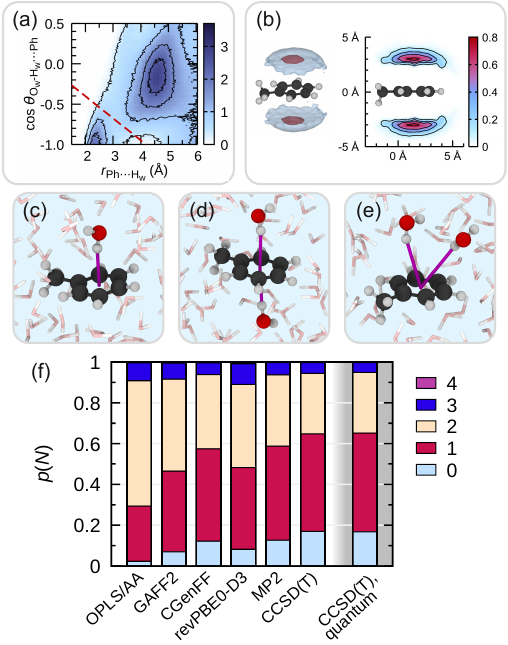}
\caption{Hydrogen bonding between the $\pi$-system of toluene and water.
(a) Joint probability density of the cosine of the angle in O$_{\rm w}$--H$_{\rm w} \cdots$Ph triplets ($\theta_{\rm O_{\rm w}-H_{\rm w}\cdots Ph}$), and the Ph$\cdots$H$_{\rm w}$ distance, from the CCSD(T) MLMD simulation.
Ph denotes the center of mass of the carbon atoms in the phenyl ring of toluene.
The joint probability density is normalized w.r.t. the probability of a random distribution.  
Contour lines are drawn at intervals of 0.7 starting at 0.3.
The red dashed line is the separatrix that defines H-bonded configurations; it connects the minimum along the $x$-axis (linear configurations) and the saddle point between the maximum in the bottom left and the rest of the distribution.
(b) Isosurfaces showing the location of O$_{\rm w}$ in water molecules that donate on average 0.15 (transparent blue) and 0.6 (opaque red) H-bonds to toluene, computed from the CCSD(T) GRACE simulation. The right figure shows a slice through the three-dimensional distribution on the left, with contour lines drawn at intervals of 0.2 starting at 0.1.
The three-dimensional distribution was calculated using voxels of side length 0.2~{\AA}, and Gaussian smoothing with $\sigma = 0.16$~{\AA} in all three dimensions was applied.
(c--e) Snapshots of water molecules donating H-bonds to the $\pi$-system of toluene.
(f) Stacked bar plot showing the probability $p$ of $N$ H-bonds being donated to the $\pi$-system of toluene, for the different FFMD and MLMD simulations.
}
\label{fig:hbonds}
\end{figure}

In all simulations, including from the CCSD(T) GRACE interaction potential, 
we find distinct H-bonded configurations near toluene, with a free energy barrier to the rest of the distribution of water near toluene~(Fig.~\ref{fig:hbonds}(a), Section~S9 in the SI).
H-bonds can be accepted on both sides of the toluene molecule, and additionally the $\pi$-orbitals (which are not explicitly modeled in our simulations but which contribute to the toluene--water interactions that are mapped to the nuclear positions with the GRACE potentials) occupy a large enough space that even two water molecules can donate an H-bond to the aromatic system, at least transiently (Fig.~\ref{fig:hbonds}(c--e)).
From the CCSD(T) GRACE simulation, we find that configurations with one accepted \mbox{H-bond} are the most probable, making up approximately 
half of all
configurations, but
depending on the interatomic potential, we find a preference for 
much
different 
numbers of $\pi$-hydrogen bonds 
(Fig.~\ref{fig:hbonds}(f)).
We 
demonstrate
that the
number of accepted H-bonds by toluene is essentially independent of the inclusion of nuclear quantum effects.
In stark contrast, it 
depends sensitively on the particular choice of interaction potential,
which imprints different ``hydrophilicities" of the aromatic ring.

For the fraction of H-bonded configurations, CGenFF results happen to be close to the CCSD(T) 
benchmark,
with a root-mean square deviation (RMSD) of 3~\%.
In contrast, the 
OPLS/AA and GAFF2 force fields 
severely
overestimate the tendency of toluene to accept H-bonds from water
via $\pi$-interactions, 
with especially configurations with two H-bonds being overrepresented; OPLS/AA is seen
to deviate most from the CCSD(T) reference. 
The revPBE0-D3 GRACE potential predictions have an RMSD of 6~\% from CCSD(T), while the MP2 GRACE potential 
achieves 
agreement within 2~\%.
Importantly, the
hybrid DFT functional overestimates 
the importance of $\pi$~solvation shells
with two accepted H-bonds, 
but
underestimates 
at the same time solvation patterns
without 
any $\pi$-hydrogen bonds 
(Fig.~\ref{fig:hbonds}(f)). 
This leads to a misrepresentation of the ``hydrophilic" toluene-water H-bonding pattern 
that is mediated by the non-covalent $\pi$-interactions due the aromatic ring.

Overall, the number of $\pi$-hydrogen bonds formed between water and toluene is 
strongly
dependent on the interaction potential. 
That is challenging in particular for the description of biomolecular systems, which are typically modeled using force fields, and where solvent--solute H-bonds are of determining importance in for example ligand binding and protein folding
in view of the subtle balance of ``hydrophilic" versus ``hydrophobic" solvation. 
We demonstrate that predictive simulations of such systems are not possible using 
these representatives of
three commonly employed force field
families, 
and even hybrid DFT predictions 
fall short in providing quantitative agreement with CCSD(T).

\subsection*{Solvation free energy barriers due to $\pi$-water interactions} 
\begin{figure}
\includegraphics[width=8.6cm]{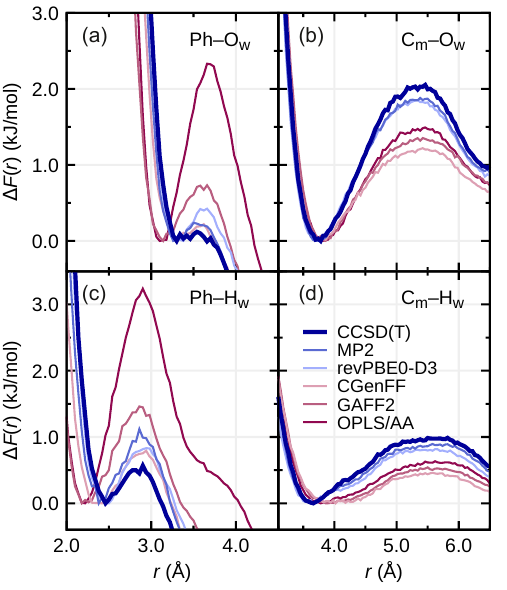}
\caption{Free energy profiles along the (a) \mbox{Ph--O$_{\rm w}$}, (b) \mbox{C$_{\rm m}$--O$_{\rm w}$}, (c) \mbox{Ph--H$_{\rm w}$} and (d) \mbox{C$_{\rm m}$--H$_{\rm w}$} distances of aqueous toluene,
from simulations with the CCSD(T), MP2, and revPBE0-D3 GRACE interaction potentials, and CGenFF, GAFF2, and OPLS/AA force fields.
The radial bin size is 0.04~{\AA}, and curves are not smoothed.
Refer to insets of Fig.~\ref{fig:rdf} for definitions of the relevant coordinates.
}
\label{fig:pmf}
\end{figure}

To shed light on the stability of the
$\pi$-hydrogen bonds
and thus exchange of ``hydrophilic water", 
we assess the solvation free energy barriers predicted by the different interaction potentials (Fig.~\ref{fig:pmf}).
The CCSD(T)  
reference
predicts a free energy barrier of approximately 0.57~\mbox{kJ/mol} in the \mbox{Ph--H$_{\rm w}$} coordinate, but there is essentially no barrier in the \mbox{Ph--O$_{\rm w}$} coordinate.
In other words, there is no barrier to breaking of the H-bond through diffusion of the water molecule, but only through rotation of the \mbox{O$_{\rm w}$--H$_{\rm w}$} bond
w.r.t.~the plane of the phenyl ring. 
Based on the solvation properties of aqueous toluene (Fig.~\ref{fig:solvationshell}(a--f)), we speculate that exchange of water molecules near the $\pi$-orbitals occurs primarily by molecules moving from or into one of the six lobes of higher water density that surround the phenyl ring.
It happens that
the CGenFF force field 
yields
the best free energy profiles of the force fields compared to CCSD(T), 
in addition to it closely reproducing 
the H-bonding pattern
(Fig.~\ref{fig:hbonds}(f)).
The OPLS/AA and GAFF2 force fields 
grossly overestimate the barrier to H-bond breaking in both coordinates, which aligns with the finding that these two force fields severely 
overpredict 
hydrophilic solvation of the phenyl ring by one and two water molecules (Fig.~\ref{fig:hbonds}(f)).

The two electronic structure methods under consideration in addition to CCSD(T), namely revPBE0-D3 and MP2, 
predict much too large free energy barriers to breaking of the $\pi$-hydrogen bonds relative to CCSD(T), by 
roughly a factor of 
two.
In the Ph--H$_{\rm w}$ coordinate, the MP2 GRACE potential has a free energy barrier of approximately 1.12~\mbox{kJ/mol}, while the revPBE0-D3 GRACE potential 
yields
a barrier of $\approx$~0.83~\mbox{kJ/mol}.
The trend is reversed for the Ph--O$_{\rm w}$ coordinate, where the revPBE0-D3 GRACE potential has a larger barrier
than MP2, whereas CCSD(T) provides 
an essentially barrierless
pathway instead. 
Thus, in
both coordinates, both methods 
markedly
overestimate the barrier relative to CCSD(T),
impacting H-bond dynamics and possibly H-bond breaking mechanisms.
In conclusion, 
predictions of solvation shell water dynamics near $\pi$-systems of aromatic molecules 
and exchange of hydrophilic water in contact with aromatic rings
are 
strongly method-dependent, 
as our data 
show
w.r.t. CCSD(T).

The trend in the free energy barriers 
for hydrophobic water exchange
(along the \mbox{C$_{\rm m}$--O$_{\rm w}$} and \mbox{C$_{\rm m}$--H$_{\rm w}$} coordinates
involving the methyl group)
is different to the trend
for exchange of $\pi$-hydrogen bonded ``hydrophilic" water
(i.e., in the coordinates involving the Ph center)
(Fig.~\ref{fig:pmf}(b,d)): Now,
the CCSD(T) GRACE potential has the largest barriers, and the CGenFF force field predicts the smallest 
ones. 
The picture that is painted suggests that these rather simple yet popular force fields that rely on isotropic interatomic Lennard-Jones and electrostatic interactions do not capture the anisotropy in interaction strength that in reality exists in molecules with delocalized electronic orbitals, such as the aromatic system in toluene.
The force fields must therefore strike a balance between providing enough attractive interactions between solute and solvent so that water molecules that only ``see" a few atoms of the solute (here, water molecules in the plane of the phenyl ring of toluene and near the methyl group) are sufficiently localized, while
water molecules that are directly on top of a \mbox{$\pi$-system} (here, water molecules located directly above the center of the phenyl ring of toluene) and thus interact with many solute atoms at once are not too strongly attracted.
For that reason, it is challenging if not impossible to accurately describe all types of solvating water molecules near organic solutes with various kinds of solute--solvent interactions using such simple force fields, 
which therefore distort the interactions of aromatic molecules with solvating water.
Systematic improvements can be achieved by introducing anisotropic descriptions~\cite{Stone1988Some}.
Still, 
even dispersion-corrected hybrid DFT, namely 
the revPBE0-D3 GRACE potential, yields a grossly unbalanced description
of hydrophilic versus hydrophobic interactions of aromatic compounds in water: 
While hydrophobic solvation 
of the methyl group is rather well-described compared to CCSD(T), the DFT functional evidently fails to get $\pi$-hydrogen bonding of the aromatic group
and thus its hydrophilic solvation right.

\section*{Discussion and Conclusions}

In this work, we 
generated a machine learning interaction potential (MLIP) based on the graph atomic cluster expansion (GRACE) 
that reproduces quantum-chemical CCSD(T) accuracy for 
simulations of 
toluene in bulk liquid water at ambient conditions.
This specific compound has been selected 
as a prototypical 
representative of molecular species that
offer both aliphatic and aromatic regions that are in contact with
what is often called ``hydrophobic" and ``hydrophilic" solvation water, respectively. 
We demonstrate that upfitting from PBE DFT or MP2 GRACE potentials reproduces
CCSD(T) energies and forces for bulk aqueous solution of toluene.
With our highly accurate 
interaction potential, we 
determined the complex 
solvation structure of aqueous toluene
with a focus on the antagonistic
hydrophobic interactions and $\pi$-hydrogen bonding of
the respective aliphatic and aromatic functional groups of
toluene and water.

Comparing now these well-converged CCSD(T) 
benchmark data~-- unavailable before~-- to results 
obtained from hybrid density functional
and 
several
force field-based simulations,
namely \mbox{revPBE0-D3}, CGenFF, GAFF2, and OPLS/AA, 
we find that none of these widely-used methods is able to correctly describe
the essential hydrophobic interactions 
and non-covalent $\pi$-bonding with water
that govern solvation of aliphatic and aromatic compounds. 
Importantly, the specific failures of the different interaction potentials
do
not follow any trend but
are 
overall erratic and incoherent, 
yielding close matches with CCSD(T) for some properties
but glaring deviations for others.
For instance,
the broadly accepted dispersion-corrected hybrid \mbox{revPBE0-D3} functional 
overestimates free energy barriers to breaking of $\pi$-hydrogen bonds 
by close to a factor of two, implying too slow H-bond dynamics
and suppressed hydrophilic water exchange near the $\pi$-electron system of the aromatic ring. 
Further, even the solvation pattern of the aromatic $\pi$-system 
is ill-described: The importance of accepting two H-bonds is
overestimated relative to the CCSD(T) reference while 
solvation by 
fewer
water molecules is underpredicted.
At the same time the hydrophobic solvation shell with 
its straddling water arrangement around the methyl group is rather well-represented
by the hybrid functional.
Overall, we 
reveal
an
inconsistent
accuracy
that leads 
to an incongruous description of
hydrophobic versus hydrophilic solvation of 
aliphatic and aromatic functional groups of organics in 
aqueous environments, as predicted by this robust DFT functional
that is widely used in water research.

Assessing next the quality of force fields that represent
three distinct families confirms the DFT picture:
No consistent match with respect to CCSD(T) data is found for a single case.
For $\pi$-hydrogen bonds, CGenFF shows remarkable agreement with the benchmark,
while OPLS/AA and GAFF2 unacceptably overestimate the corresponding free energy barrier
for water exchange, 
and, accordingly, also overestimate the total number of hydrophilic water
molecules that are H-bonded to the aromatic ring.
In contrast, CGenFF is the worst force field when it comes to
reproducing the hydrophobic solvation shell around the aliphatic group,
underestimating the exchange free energy for hydrophobic water by far, 
whereas, for this property, OPLS/AA and GAFF2 come closer to CCSD(T). 
We therefore must conclude that these force field families 
all fail to provide a consistent description of the key interactions
of this generic organic compound with water,
namely hydrophobic interactions and intermolecular $\pi$-contacts. 
Based on these insights, one might question the accuracy and predictive power
of
biomolecular simulations 
where hydrophobic interfaces and local H-bonds are of determining importance,
for example
in biomolecular recognition-based phenomena which hinge 
on the very balance of hydrophilic versus hydrophobic solvation.

Non-polarizable 
force fields relying on isotropic Van der Waals and electrostatic 
intermolecular
interactions, such as the 
representatives examined 
here, 
do
not capture the anisotropy of aromatic molecules 
interacting 
with water,
as judged by one-to-one comparisons to now-available CCSD(T) benchmark data
for the condensed phase.
Fundamentally, if the hydrophobic interactions are parameterized correctly by such a force field, then hydrophilic interactions with $\pi$-systems are severely overestimated, 
and \emph{vice versa}.
We observe exactly this trend in our 
simulations that scrutinize three complementary such force fields
that are 
very ubiquitously 
used
in large-scale investigations 
of biomolecular processes.
Thanks to our accurate CCSD(T) reference, we show that the 
water density near the $\pi$-orbital system of 
the aromatic ring
is decreasing 
from OPLS/AA to GAFF2 to CGenFF,
with the CGenFF 
force field
providing results closest to CCSD(T);
the first solvation shell of the methyl group follows 
the same trend, but in this case OPLS/AA provides results closest to CCSD(T).
Improving 
such non-polarizable
force fields for one kind of intermolecular interaction of the anisotropic molecule 
evidently
deteriorates 
their 
description for other non-covalent interactions
in a significant way. 
What is more, none of the three force field families 
capture
the 
decisive straddling
orientations of hydrophobic water 
correctly, further providing evidence that describing both hydrophilic and hydrophobic 
water around multifunctional organic compounds 
leads to irreconcilable errors.
In sum,
the errors of 
such 
force fields preclude their use in accurate simulations of molecular 
species in which hydrophobic groups as well as aromatic rings are exposed to water.

Beyond the specific case, the distinct solvation properties
of hydrophobic versus aromatic functional groups are 
important
in proteins, nucleic acids, and small-molecule ligands.
Overbinding of water molecules at $\pi$-orbitals could affect 
binding free energies, altering conformational equilibria, and leads 
to slower water exchange near aromatic groups, changing for 
example ligand residence times
or water-mediated $\pi$-stacking. 
Understructuring of the water solvation shell near 
hydrophobic 
groups on the other hand could alter hydrophobic aggregation
and collapse,
crucial for the correct description of protein folding and membrane formation.
An accurate description of biochemical compounds in water thus requires both aromatic and hydrophobic water contacts to be correctly represented. 
Only interaction potentials that recover the CCSD(T) balance between $\pi$-interactions and hydrophobic interactions can be considered predictive for biomolecular simulations.

Given the evidence that three important force field families fail
to reliably describe 
both hydrophobic and hydrophilic contacts 
already at the level of chemically simple 
aqueous organic molecules, 
it is suggested that fundamental improvement 
of biomolecular force fields 
can only be offered by polarizable force fields that incorporate anisotropy~\cite{Stone1988Some, Lemkul2016Empirical, Zhang2018AMOEBA, Janicki2023Development} or more flexible models, including MLIPs, fitted to reproduce correlated wave function calculations~\cite{Zhou2025Water}.
The training set construction workflow that we present in this work, together with the GRACE machine learning methodology, represents a general, practical, and transferable approach to training MLIPs for aqueous solutes with CCSD(T) accuracy.
For 
those
extended molecules where long-range charge transfer is expected to be important, explicit inclusion of electrostatic interactions in the machine learning interatomic potential will be needed~\cite{Ko2021Fourth-Generation, Rinaldi2025Charge}.
Our
strategy will enable 
quantitative insights into the solvation 
structural dynamics
of 
chemical and biomolecular species
from environmentally to biologically relevant settings.
In this way, we 
systematically
close the accuracy gap between state-of-the-art electronic structure
calculations
and molecular simulation for aromatic solvation, and provide a rigorous 
benchmark for force fields and 
DFT-based 
simulations of aqueous 
solutes 
that include aromatic functionalities, from residues in proteins to 
nucleic acids to polycyclic aromatic hydrocarbons.

\section*{Methods}

\subsection*{Machine learning interatomic potential training}

A detailed description of the training and validation of the CCSD(T) GRACE interaction potential for aqueous toluene, as well as the revPBE0-D3 and MP2 GRACE interaction potentials for which we report results, is included in the 
SI.
To provide the basis for our
CCSD(T) upfitting
strategy that is exclusively based on using finite cluster data, 
in the SI
we demonstrate 
first, 
using more economical DFT calculations 
in a ``mock training procedure"
detailed in
Section~S1,
that our methodology results in an MLIP that 
perfectly
reproduces structural properties for aqueous toluene from 
periodic 
reference simulations.
We achieved a GRACE potential for
dispersion-corrected hybrid DFT
revPBE0-D3~\cite{Zhang1998Comment,Adamo1999Toward,Goerigk2011Thorough, Grimme2010Consistent}
calculations
through upfitting of a GRACE potential trained to PBE DFT~\cite{Perdew1996Generalized} energies and forces of 
finite homo- and hetero-clusters of water and aqueous toluene containing 1--64 molecules,
using exclusively \mbox{revPBE0-D3} energies (see Section~S4).
Here, 
we selected \mbox{revPBE0-D3} since it is a widely
used functional to reliably describe H-bonding and aqueous systems. 
Given the purpose, we chose the GGA-quality PBE functional without any dispersion correction
as the starting 
point 
to validate our upfitting procedure
because it provides a very different description of water and aqueous solutions than
revPBE0-D3,
thus demonstrating that 
our upfitting approach
correctly moves the 
simple PBE
GRACE potential to 
the level of dispersion-corrected 
hybrid DFT accuracy.
Importantly, the 
finite cluster
training structures consist of atomic configurations extracted from periodic simulations of neat water and aqueous toluene.

\begin{figure}
\includegraphics[width=8.6cm]{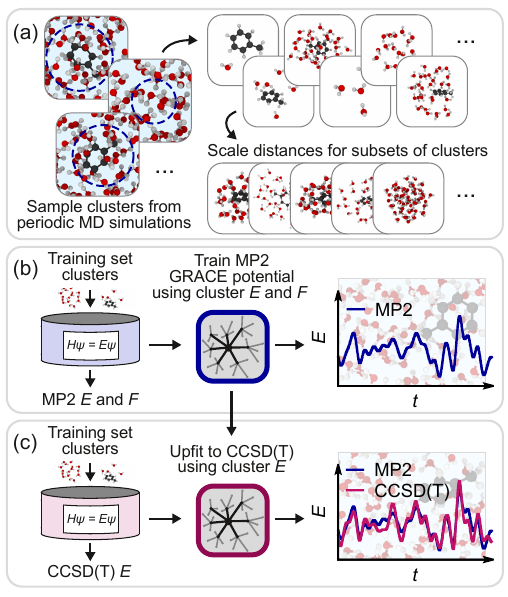}
\caption{Scheme illustrating the training procedure of the CCSD(T) GRACE 
interaction potential 
for aqueous toluene.
(a) Training set structures generation scheme
based on homo- and hetero-clusters of toluene and water molecules.
(b) Training procedure of the MP2 GRACE interaction potential, which used MP2 energies and forces computed for the training set structures.
(c) Upfitting procedure of the CCSD(T) GRACE interaction potential, which 
exclusively
used CCSD(T) energies for the training set structures. 
}
\label{fig:scheme_extended}
\end{figure}

To subsequently achieve the CCSD(T) quality potential for aqueous toluene, we first trained a GRACE interaction potential to MP2 accuracy using a training set of finite clusters labeled with MP2 energies and forces.
We then constructed a smaller training set of such finite clusters and performed CCSD(T) energy calculations for those.
Starting from the MP2 GRACE potential, we trained a CCSD(T) GRACE potential using upfitting, employing only energies of finite cluster data;
Fig.~\ref{fig:scheme_extended} summarizes the procedure.
In the SI, we report end-to-end testing results that verify the quality of our trained potentials.  
Notably,
we also demonstrate 
therein
that the forces provided by the upfitted CCSD(T) potential do not depend on the quality of the starting potential, here MP2.
Nevertheless, here
our final procedure to train GRACE potentials for aqueous solutions at CCSD(T) accuracy does not make use of DFT data.
 
For training we employed the \texttt{gracemaker} code~\cite{Bochkarev2024Graph, Lysogorskiy2026Graph, gracemaker}.
The GRACE models use a 2-layer architecture with the pre-set ``large" complexity, and a 5~{\AA} cutoff.
We used a 9:1 train:test split during training, and the final model is that with the smallest test loss after 1500 optimization epochs.

\subsection*{Reference calculations}

Reference calculations were carried out with \texttt{ORCA}~\cite{Neese2012ORCA, Neese2025ORCA}
for correlated wave function calculations, and
\texttt{CP2K}~\cite{Hutter2014CP2K, 
Kuhne2020CP2K, Iannuzzi2026CP2K, CP2K} for DFT.
Correlated wave function calculations that formed the basis for our GRACE potentials used the 
DLPNO-CCSD(T)~\cite{Riplinger2016Sparse, Neese2019Chemistry}
and 
DLPNO-MP2~\cite{Pinski2015Sparse, Pinski2019Analytical}
methodology,
as comprehensively validated 
for the present application 
in Sections~S5 and S3, respectively.
The DLPNO-CCSD(T) calculations used
the def2-TZVPP orbital basis for the
DLPNO-CCSD(T) calculation, and def2-QZVPP basis for the Hartree-Fock energy.

\subsection*{Simulations}
We performed 3~ns long molecular dynamics simulations of a toluene molecule in a 
periodic
simulation cell with 256 H$_2$O in the NVT ensemble, at 298.15~K and with volume corresponding to the experimental density of the solution~\cite{Sawamura2001Effects} at 1~atm.
To obtain the interatomic forces underlying the reported simulations, 
we used in the first place our GRACE interaction potentials based on 
DLPNO-CCSD(T)
and 
DLPNO-MP2
correlated wave function calculations,
as well as the revPBE0-D3 GRACE interaction potential.
Finally, three 
widely used biomolecular force fields were used in simulations, namely 
GAFF2~\cite{Wang2004Development} 
with TIP3P water~\cite{Jorgensen1983Comparison},
CGenFF~\cite{Vanommeslaeghe2010CHARMM} with CHARMM-modified TIP3P water~\cite{MacKerell1998All-Atom},
and
OPLS/AA~\cite{Jorgensen1996Development, Jorgensen2005Potential} with TIP4P water~\cite{Jorgensen1983Comparison, Jorgensen1985Temperature}, 
to represent three complementary force field families with extensive optimization histories.
To assess nuclear quantum effects we 
carried out a 500~ps 
long path integral simulation~\cite{MarxHutter2009} of aqueous toluene using the CCSD(T) GRACE potential at the same conditions with the PILE thermostat~\cite{Ceriotti2010Efficient}, with a Trotter discretization of 32.
We used \texttt{GROMACS}~\cite{Abraham2015Gromacs} for the 
FFMD
simulations, and \texttt{CP2K}~\cite{CP2K} 
with 
our 
interface to 
\texttt{gracemaker}
and its path integral module~\cite{Brieuc2020Converged, Iannuzzi2026CP2K} for the 
MLMD
simulations.
More details are provided in Section~S2.

\section{Data Availability}
All data needed
to evaluate the conclusions in the paper are present in
the paper and/or the Supporting Information.

\section{Supporting Information}
Description of the GRACE training procedures, details of the electronic structure calculations, validation of the DLPNO-CCSD(T) calculations, end-to-end validation of the GRACE interatomic potentials, details of the simulations, assessments of finite size effects and nuclear quantum effects, spatial distribution functions and H-bond donor distributions from all simulations.

\section{Acknowledgments}
We are grateful to Frank Neese for most valuable discussions and suggestions
on DLPNO-CCSD(T) calculations. 
Funded by the Deutsche Forschungsgemeinschaft (DFG, German Research Foundation) under Germany's Excellence Strategy~-- EXC~2033~-- 390677874~-- RESOLV.
All computations have been carried out locally at HPC@ZEMOS, HPC-RESOLV, and BOVILAB@RUB.

\bibliography{refs}

\end{document}